# Major Revision, Minor Accountability: LLMs as Reviewers of Published Articles in Top-Tier Journals

Bastien Le Guellec[1,*], Su-Hwan Kim[2], Raphael Bentegeac[1] and Aghiles Hamroun[1]

[1]CHU Lille, Lille, France
[2]Technical University of Munich (TUM), Munich, Germany

*Corresponding author: bastien.leguellec@chu-lille.fr

**Abstract**

With the growth of scientific output, the peer review system is approaching a breaking point: reviewers are increasingly hard to secure, and turnaround times lengthen accordingly. LLM-assisted or LLM-generated peer review has been proposed as a remedy, but how such models would treat manuscripts already published in top-tier journals – the closest thing the current system has to a gold standard of scientific validity – is not known. Here we assembled a corpus of 200 articles, each released under a CC BY 4.0 license and published in 2026, from 20 general-science, life-science, and clinical-medicine journals including *Nature*, *Science*, and *Cell*, and asked four current LLMs – two proprietary, two open-weight – to produce independent peer-review reports of each one under a blinded protocol. Their decisions were overwhelmingly harsh and starkly inconsistent with each other: 605 of 800 reviews (75.6%) recommended major revision or rejection, but the rate at which any single model did so ranged from 17.5% to 100% ($p < 0.001$) – disagreement on manuscripts that had, by construction, already survived real editorial review at the journals in question. When the same manuscripts were doctored with one deliberately planted, field-specific methodological or statistical flaw, detection rates varied from 17.0% to 60.5% across models ($p < 0.001$). Recommending revision for papers that had already passed editorial review (low specificity) while missing flaws deliberately placed in front of them (low sensitivity), these models produce output that should be read for what it is – an uneven, framing- and model-dependent signal.

## 1 Introduction

Submission volumes have grown faster than the supply of reviewers willing and able to evaluate them [1, 2, 3]. Editors report escalating difficulty securing the two or three qualified reviewers a typical manuscript requires, and turnaround times have lengthened accordingly [2, 4].

Part of this growth stems from career-evaluation systems that continue to reward the number of publications (*publish or perish*) [1]. The sharp acceleration observed since 2024, however, is linked to the development of generative AI tools [5, 6]: many steps in writing a scientific article can now be assisted, or even automated. Growth has been particularly rapid in fields that do not depend on experiments in the physical world [7], to the point that editorial policies are being revisited [8]. One proposed remedy is to limit the number of submissions allowed per author [9].

Paradoxically, the very tools behind this rise in submissions can also be turned toward addressing it. Many tasks in the reviewing process can now be assisted or automated: summarizing the manuscript, drafting an initial pass at the review, or flagging obvious problems before the reviewer commits their own attention. A 2025 Frontiers Media survey found that 53% of reviewers were already using AI for peer review tasks [2]. Journals and conferences themselves lead this adoption [10]: every main-track submission to AAAI-26 (22,977 papers) already received an AI-generated review as part of a field deployment [11], and NEJM AI now offers an accelerated review track that combines LLM analysis with human editorial judgment [12].

Despite the growing reach of LLMs in reviewing, there has been comparatively little systematic evaluation of whether LLMs are actually good at the core task being delegated to them. A small empirical literature bears on the question: in an expert assessment of reviews of 82 published Nature-family papers, a reviewing agent built on GPT-5.2 scored above each paper's top-rated human reviewer on a composite of correctness, significance, and sufficiency of evidence, but AI reviewers overlapped with each other far more than human reviewers did (21% versus 3%) [13]. Howell and colleagues found that LLM reviewers are susceptible to bias, penalizing lower-prestige author affiliations [14]. They have also been shown to be vulnerable to cosmetic rewriting, or *paper laundering* [15]. Interestingly, AI-assisted reviews at ICLR systematically raised acceptance probability near the decision threshold, pointing towards indecision [16].

However, no study has investigated how LLMs would review papers already published in top-tier journals. These papers are the closest thing to a collective ground truth of what deserves to be published (see Discussion for a contextualization of this claim). By measuring how many of them would receive a favorable recommendation, we assess these models' specificity: their ability to recognize a sound paper as such. In addition, by injecting errors into the manuscripts, we assess their sensitivity: their ability to detect a flaw [13, 17].

## 2 Methods

### 2.1 Manuscript corpus

We assembled a corpus of 200 recently published research and review articles (ten per journal) from 20 peer-reviewed journals spanning general-science, life-science, and clinical-medicine venues, from broad-acceptance mega-journals (*Scientific Reports*) to flagship venues (*Nature*, *Science*, *Cell*): *Nature*, *Nature Medicine*, *Nature Communications*, *Nature Genetics*, *Nature Materials*, *Scientific Reports*, *Science*, *Cell*, *Cell Reports Medicine*, PNAS, *eLife*, *PLOS Biology*, *PLOS Medicine*, *Circulation*, the *European Heart Journal*, the *Journal of Experimental Medicine*, *JAMA Internal Medicine*, *JAMA Oncology*, *JAMA Psychiatry*, and the *Lancet Global Health*. Every included article met four criteria: publication in 2026; a CC BY 4.0 license; a body of at least 2,000 words, excluding short-format letters; and no correction, erratum, or retraction notices.

Full text was retrieved as JATS XML: via the Springer Nature Open Access API for the five Springer-published journals, and via the Europe PMC API – which exposes license terms directly – for the remaining 15. Title, abstract, and body text (including figure/table captions and in-XML table content) were concatenated into plain text per article; rendered images and supplementary files were not included.

### 2.2 Reviewer models and review protocol

Four LLMs served as reviewers, all accessed via the OpenRouter API:

- `z-ai/glm-5.3-flash` (hereafter GLM-5.3 Flash)
- `openai/gpt-5.6-sol` (hereafter GPT-5.6 Sol)
- `google/gemini-3.7-flash` (hereafter Gemini 3.7 Flash)
- `deepseek/deepseek-v4-flash-0731` (hereafter DeepSeek v4 Flash)

Models were selected on two criteria: a competitive ranking on the OpenRouter leaderboard (top 50) and a price below $5 per million output tokens. Each manuscript was submitted to each reviewer model twice: once as a first submission, and once presented as a revision already through one round of review. The two system prompts were identical except for their opening framing sentence, named the manuscript's true target journal, and asked for the review the model would submit to the editor: a summary, major and minor concerns, and a final decision (accept, minor revision, major revision, or reject). Models were not told the manuscript had already been published. The first-submission prompt read:

```
You are a reviewer for {journal}. You will be shown the full text of a submitted
manuscript. Write the review you would hand to the Editor.

Structure your review as:

SUMMARY: <2-3 sentence summary of what the paper claims to show>

MAJOR CONCERNS:
- <substantive issue that would need to be resolved before publication, or
"None.">

MINOR CONCERNS:
```

```
- <smaller issues, wording, presentation>

Then end your response with exactly one line in this format, with no other text on
that line:

DECISION: <accept|minor_revision|major_revision|reject>
```

In the revision round, the opening sentence instead read: *"You are a reviewer for {journal}, conducting the second round of peer review. You will be shown the full text of a submitted manuscript, revised after an initial round of review."* Everything else, including the required output structure and the final decision line, was unchanged. The decision was parsed programmatically from each response. Unless stated otherwise, results refer to the first-submission round; the paired comparison between rounds is reported in §3.2. Figure 1 summarizes the corpus, the reviewer panel, and the three experiments.

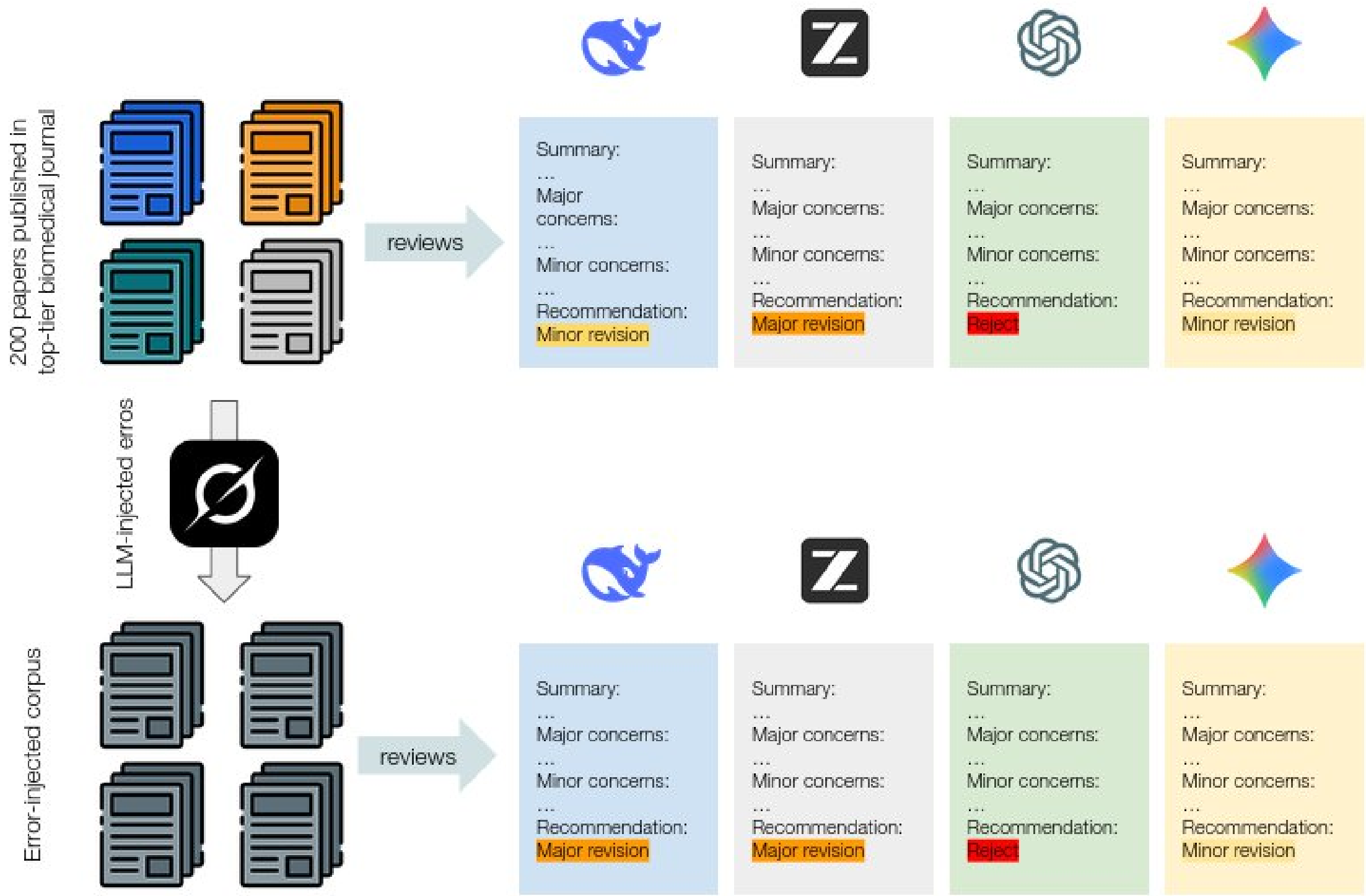

**Figure 1.** Study design. Two hundred research and review articles published in 2026 across 20 general-science, life-science and clinical-medicine journals, all CC BY 4.0, were reviewed by four LLMs under one blinded protocol that named the manuscript's true target journal and withheld the fact that it had already been published. Experiment 1 collected one review of each unmodified manuscript per model. Experiment 2 repeated the protocol after an off-panel model had planted one severe, field-specific flaw in each manuscript, with detection adjudicated by a second off-panel model. Experiment 3 presented each manuscript's four Experiment-1 reviews, relabelled and randomly ordered, to off-panel judges.

We treated the four decision categories as an ordinal severity scale (accept = 0, minor revision = 1, major revision = 2, reject = 3) for quantitative analysis, while retaining the categorical labels for descriptive reporting.

## 2.3 Error-injection experiment

To test whether reviewer models actually detect serious problems rather than producing generically plausible criticism (Experiment 2), an off-panel model (`x-ai/grok-4.6`, not part of the reviewer panel) planted one deliberate flaw into each manuscript as a targeted, verbatim string substitution rather than full-manuscript regeneration. Flaws were assigned round-robin to one of two categories: *statistical* (an analysis that does not match the data's actual structure – for example, treating clustered observations as independent) or *methodological* (an outcome assessed with undisclosed access to biasing information, or an instrument that does not isolate what is claimed). Each flaw was designed to be substantive and field-specific, severe enough to independently warrant major revision or rejection.

Each of the four reviewer models then reviewed the altered manuscript under the protocol of Experiment 1. Detection was judged by Claude Opus 5, given the manuscript's central conclusion, the planted flaw, and the review text, and scored as *yes*, *partial* or *no*; only *yes* counts as a catch below, so the 125 partial verdicts (15.6% of the 800 reviews) are counted as misses. Scoring partial verdicts as catches instead would raise detection from 60.5% to 84.0% for the best model and from 17.0% to 29.5% for the weakest; GPT-5.6 Sol and GLM-5.3 Flash would still rank first and second, and the two weakest would tie.

### 2.4 Review-quality judging experiment

The first two experiments assess a reviewer model's editorial *decision* (Experiment 1) and its ability to *detect* a specific known flaw (Experiment 2). Neither speaks directly to the quality of a review's actual content when nothing is deliberately planted – whether its concerns are substantive or superficial, and how it compares to the other models' take on the identical manuscript. To address this (Experiment 3), for each of the 200 manuscripts we took all four reviewer models' Experiment-1 reviews of the true, unmodified manuscript and had two off-panel judges score them side by side: `x-ai/grok-4.6` and `minimax/minimax-m3`.

To keep the judges' assessment blind to model identity, the four reviews for a given manuscript were relabeled "Review A," "Review B," "Review C," and "Review D" in an order randomized independently per manuscript, and the judge prompt made no reference to the reviews being AI-generated at all – they were presented exactly as four anonymous referee reports. Given the manuscript and its four relabeled reviews, the judge was asked to: (i) identify every major concern raised across all four reviews and group substantively identical concerns – even if differently phrased – into a single cluster, recording every review letter that raised it; (ii) classify each concern cluster as *relevant* (a real, well-founded issue that could legitimately affect the paper's conclusions or acceptance) or *nitpicking* (a technically valid but minor, pedantic, or already-addressed point); (iii) rate each review's overall constructiveness on a 1–5 scale, independent of how harsh or lenient its recommendation was; and (iv) name the single best review of the four and give a one-sentence justification. The full judge prompt is reproduced in Appendix A. Both judges scored all 200 manuscripts, but for six of them `minimax/minimax-m3` exhausted the default 4,000-token completion budget on its reasoning trace and returned no parseable output; those six were re-run with a 16,000-token budget, which is the only protocol difference between them and the other 194.

### 2.5 Statistical analysis

All statistical tests were computed in Python (SciPy 1.16, statsmodels 0.14). Differences in decision severity across the four reviewer models (Experiment 1) were assessed with the Kruskal-Wallis test, appropriate for an ordinal outcome without a normality assumption. Because the same 200 manuscripts were reviewed by all four models, these observations are not independent across models; as a clustered robustness check we additionally computed a Friedman test, which treats manuscript as a blocking factor. Differences across journals were assessed with a Kruskal-Wallis test, and the comparison of *Nature*, *Science* and *Cell* with *Scientific Reports* with a Mann-Whitney U test. Paired changes in the recommendation for the same manuscript and model – between the first-submission and revision framings, and between the original and the flaw-injected manuscript – were assessed with the Wilcoxon signed-rank test. Differences in detection rate across models (Experiment 2) were assessed with a $\chi^2$ test of independence on the 4-model × 2-outcome (caught/missed) contingency table, and, as a clustered robustness check, with a logistic generalized estimating equation (exchangeable working correlation, clustered on manuscript); detection by flaw category was compared with a $\chi^2$ test on the 2-category × 2-outcome table. For the review-quality judging experiment (Experiment 3), the distribution of best-review picks across the four models was assessed with a $\chi^2$ goodness-of-fit test against a uniform null (each model picked best equally often, 25%); the relationship between the number of reviewers independently raising a concern cluster and its relevance was assessed with a $\chi^2$ test of independence on the 4-agreement-level × 2-category contingency table; agreement between the two judges was summarized with Cohen's $\kappa$ for the best-review pick and a Spearman correlation for the constructiveness scores. We report descriptive means on the four-point ordinal severity scale (accept = 0 through reject = 3) alongside these rank-based tests for interpretability; all inferential tests, however, use ranks or exact paired differences rather than the interval assumption a mean difference implies. All reported p-values are two-sided unless otherwise noted, and statistical significance was set at $p < 0.05$.

## 3 Results

### 3.1 Reviewer models disagree sharply on the same manuscripts

Across the same 200 manuscripts, the four reviewer models produced markedly different distributions of recommended decisions (Figure 2): GPT-5.6 Sol recommended major revision or rejection for 200/200 manuscripts (100%), GLM-5.3 Flash for 180/200 (90.0%), Gemini 3.7 Flash for 35/200 (17.5%), and DeepSeek v4 Flash for 190/200 (95.0%) (Kruskal-Wallis $H = 465.0$, $p < 0.001$, $\varepsilon^2 = 0.58$; Friedman test blocking on manuscript, $\chi^2 = 426.5$, $p < 0.001$).

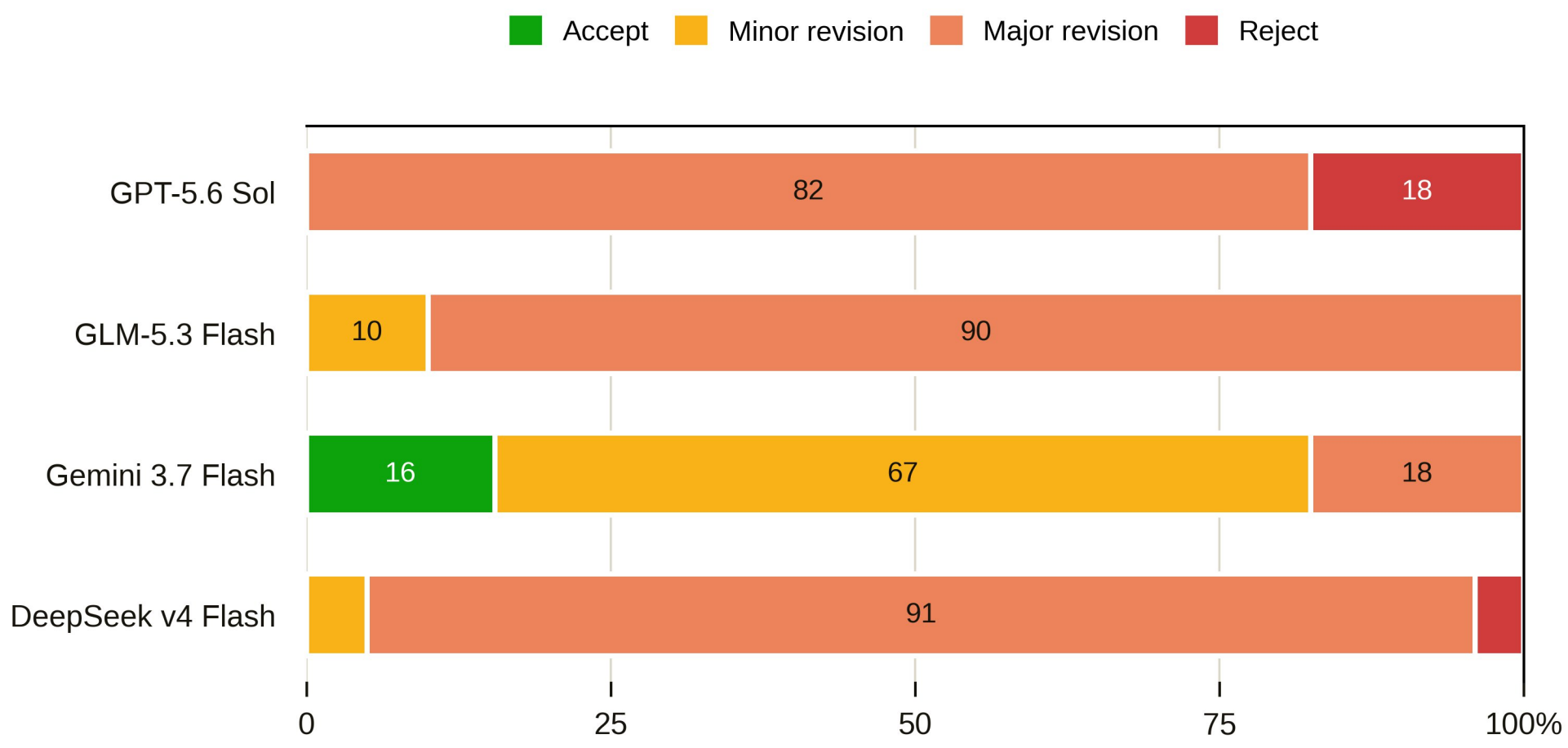

**Figure 2.** Distribution of recommended editorial decisions returned by each of the four reviewer models across the same 200 manuscripts (n = 200 reviews per model).

Considering all four models jointly (Figure 3), only 21/200 manuscripts (10.5%) received an identical decision from every reviewer – always a unanimous major-revision call, never a unanimous accept – spread across 14 of the 20 journals.

Recommendations did not differ significantly by journal (Kruskal-Wallis across 20 journals, 40 reviews each, $H = 23.0$, $p = 0.24$), nor with prestige: the mean recommendation at *Nature*, *Science*, and *Cell* (1.74 on the four-point scale, n = 120 reviews) was not significantly more lenient than at *Scientific Reports* (1.98, Mann-Whitney $p = 0.081$).

### 3.2 Presenting the same manuscript as a revision makes every model more lenient

Re-reviewing all 200 manuscripts under a prompt identical except for an opening sentence stating that the text had already been through one round of review, every model became more lenient (Figure 4). Pooled across models, mean severity fell from 1.77 to 1.26 on the four-point scale (n = 800 paired reviews; Wilcoxon signed-rank $p < 0.001$): 392/800 reviews (49.0%) moved to a milder recommendation, 402 (50.3%) did not move, and 6 (0.8%) moved harsher. The shift was significant for all four models (n = 200 pairs each) and its size tracked baseline leniency, from −0.70 for the most lenient model (Gemini 3.7 Flash, $p < 0.001$) through GLM-5.3 Flash (−0.66, $p < 0.001$) and DeepSeek v4 Flash (−0.51, $p < 0.001$) to −0.18 for the harshest (GPT-5.6 Sol, $p < 0.001$). Accept rose from 3.9% to 17.8% of reviews and minor revision from 20.5% to 39.1%, while major revision fell from 70.3% to 42.3% and outright rejection nearly disappeared (5.4% to 0.9%). For Gemini 3.7 Flash the accept rate rose from 15.5% to 68.5%, so its majority verdict on the same published manuscripts flips from minor revision to accept on one sentence of context.

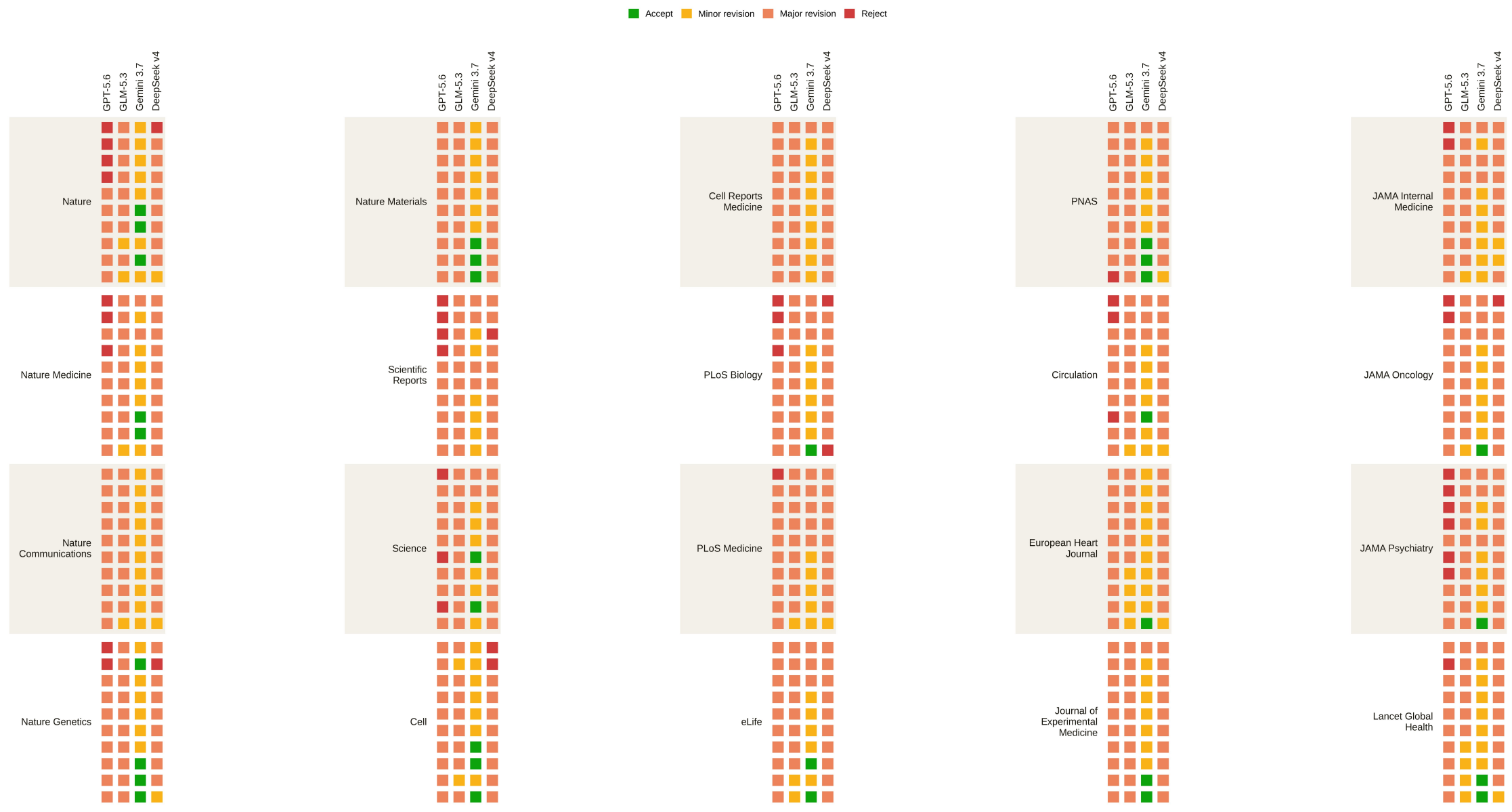

**Figure 3.** Recommended decision for every one of the 200 manuscripts (rows), by reviewer model (columns), grouped by journal. Colour encodes decision severity from accept (green) to reject (red); white rules separate journal groups.

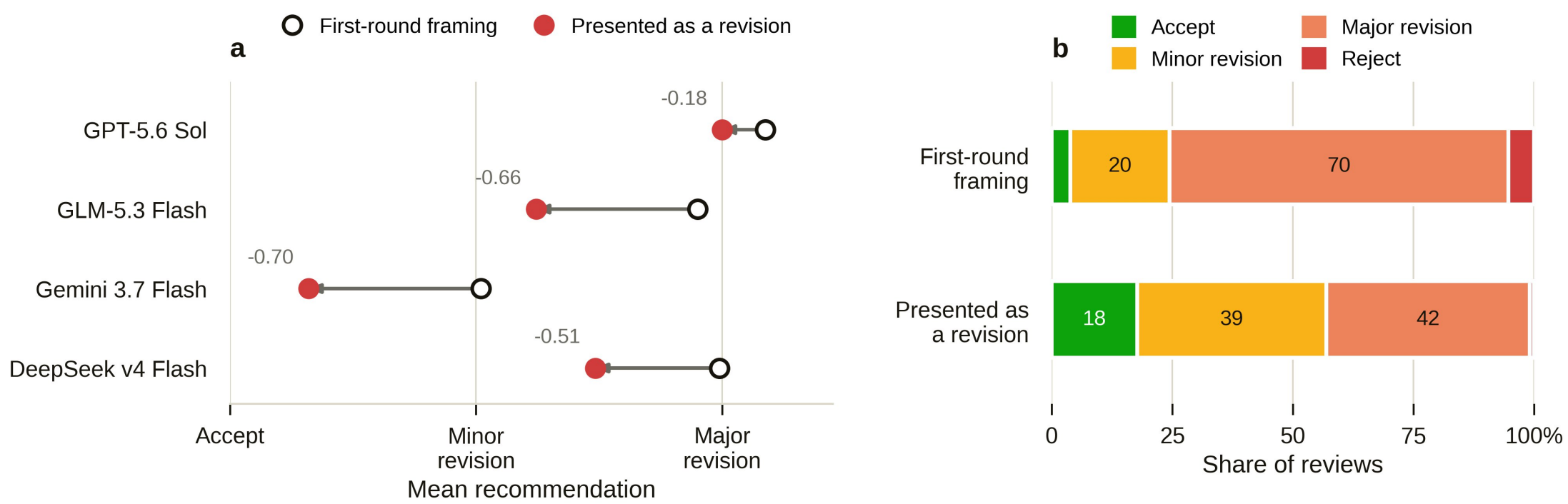

**Figure 4.** Effect of presenting the same manuscript as a revision rather than as a first submission (200 manuscripts × 4 reviewer models = 800 paired reviews). The two prompts are identical apart from an opening sentence stating that the manuscript had already been through one round of review. **a**, Mean recommendation under each framing, by model; the number above each pair is the mean paired shift, negative meaning a milder recommendation. **b**, Pooled distribution of recommended decisions under the two framings.

## 3.3 Reviewer models miss most single planted flaws, and detection rate itself varies by model

Detection of the single, deliberately planted, field-specific flaw was inconsistent across models: GPT-5.6 Sol caught 121/200 (60.5%), GLM-5.3 Flash 88/200 (44.0%), Gemini 3.7 Flash 36/200 (18.0%), and DeepSeek v4 Flash 34/200 (17.0%) ($\chi^2 = 118.4$, $p < 0.001$; Figure 5), although the two weakest could not be distinguished from each other (manuscript-clustered logistic GEE, $p = 0.76$). Detection did not differ between the two flaw categories, *statistical* (151/400, 37.8%) and *methodological* (128/400, 32.0%; $\chi^2 = 2.66$, $p = 0.10$), and GPT-5.6 Sol and GLM-5.3 Flash ranked first and second in both.

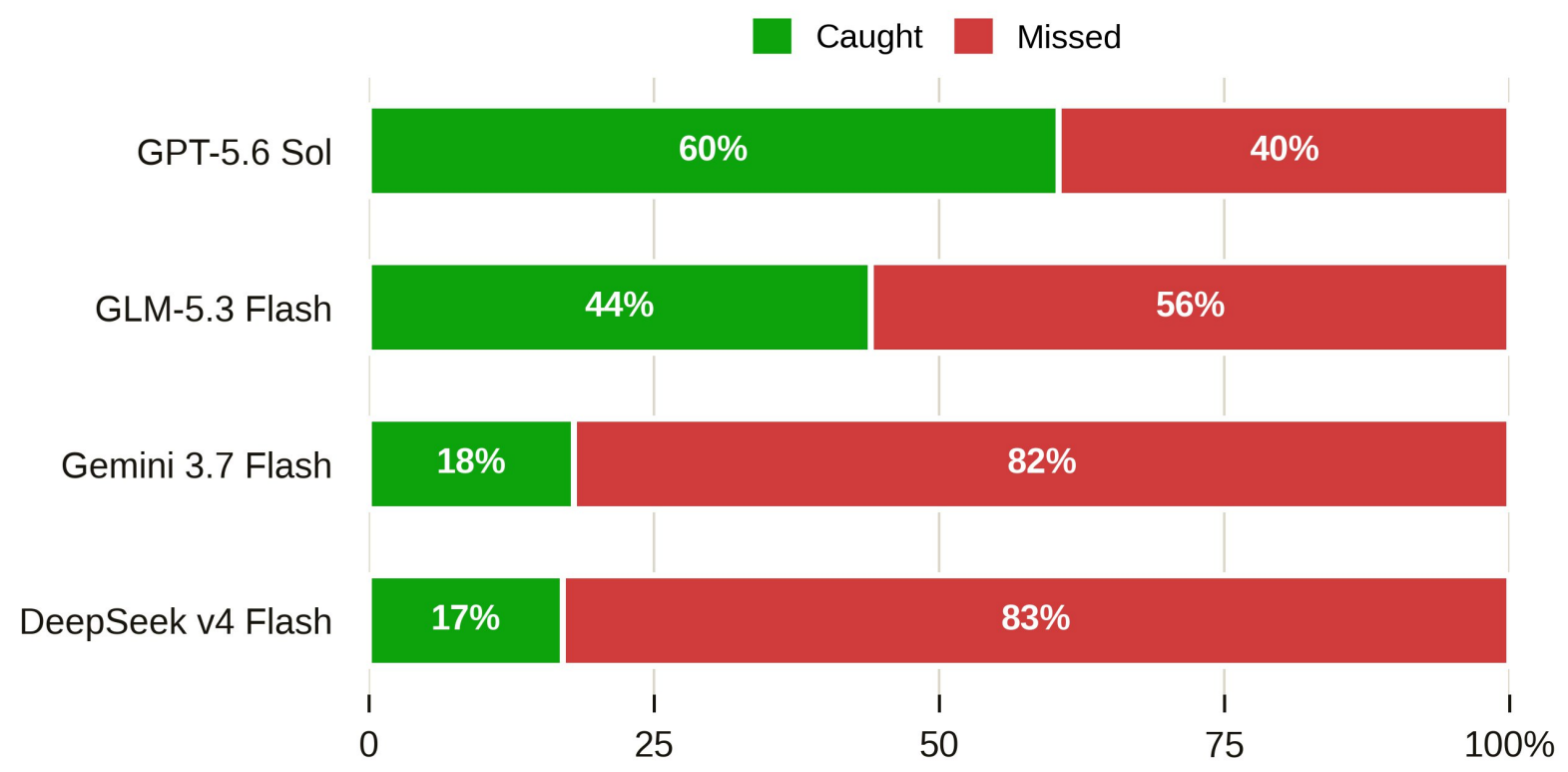

**Figure 5.** Detection rate for a single, deliberately planted, field-specific critical flaw, by reviewer model ($n = 200$ manuscripts per model). The two flaw categories are pooled here, detection having not differed between them.

Planting a verifiable flaw nonetheless barely moved the recommendation itself. Paired against the same model's review of the unaltered manuscript, mean severity rose by 0.02 of a category across all 800 paired reviews (GPT-5.6 Sol +0.02, GLM-5.3 Flash 0.00, Gemini 3.7 Flash +0.04, DeepSeek v4 Flash +0.03; Wilcoxon signed-rank $p > 0.35$ for every model, $n = 200$ pairs each), and the share of manuscripts recommended for major revision or rejection changed by at most four manuscripts in any model. Restricting to the reviews in which the judge confirmed the planted flaw had been identified, severity rose by 0.07 of a category ($n = 279$ reviews, Wilcoxon signed-rank $p = 0.006$) and the recommendation was unchanged for 84.2% of them, against no significant movement where the flaw was missed (−0.02, $p = 0.36$).

### 3.4 One model's reviews are judged best almost universally, and cross-model agreement tracks genuine relevance

Both judges named the same winning review. GPT-5.6 Sol's was picked best of the four in 94.5% of manuscripts by `x-ai/grok-4.6` (189/200; $\chi^2$ goodness-of-fit against uniform, $\chi^2 = 516.8$, $p < 0.001$) and in 75.5% by `minimax/minimax-m3` (151/200; $\chi^2 = 304.0$, $p < 0.001$); GLM-5.3 Flash took the remainder, and neither Gemini 3.7 Flash's nor DeepSeek v4 Flash's review was ever named best by either judge (Figure 6a). GPT-5.6 Sol also raised the most relevant concerns, was rated most constructive, and had the lowest nitpicking rate under both judges, which ordered the four models identically on constructiveness (item-level Spearman $\rho = 0.83$ across 800 review-level scores; Figure 6b–d). The two judges picked the same best review for 79.0% of the 200 manuscripts (Cohen's $\kappa = 0.23$, deflated by the concentration of wins in one model rather than by disagreement about which model wins). Both also treated cross-model agreement as a relevance signal: concerns raised by a single reviewer were judged nitpicking far more often than concerns raised by all four (`x-ai/grok-4.6`: 11.9% versus 1.2% of 2,175 concern clusters, $\chi^2 = 122.2$, $p < 0.001$; `minimax/minimax-m3`: 31.9% versus 20.5% of 3,112 clusters, $\chi^2 = 88.6$, $p < 0.001$), an overlap that is structured by model (Figure 7): Gemini 3.7 Flash's small pool of concerns is almost entirely a subset of the other three models', which overlap with each other more evenly.

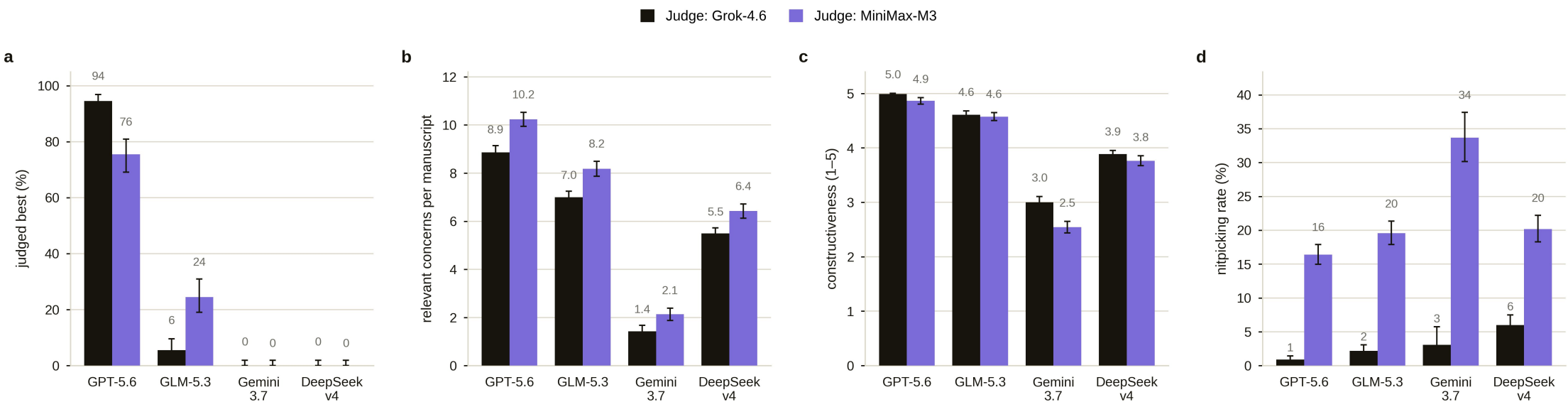

**Figure 6.** Blind quality judging of each manuscript's four reviews under two independent judge models – `x-ai/grok-4.6` and `minimax/minimax-m3` – on all 200 manuscripts, each judge seeing the four reviews anonymised and in randomised order. **a**, Share of manuscripts for which this model's review was named best of the four. **b**, Mean number of relevant concerns raised per manuscript. **c**, Mean judge-rated constructiveness, 1–5. **d**, Share of this model's raised concerns judged nitpicking rather than relevant. Error bars are 95% confidence intervals.

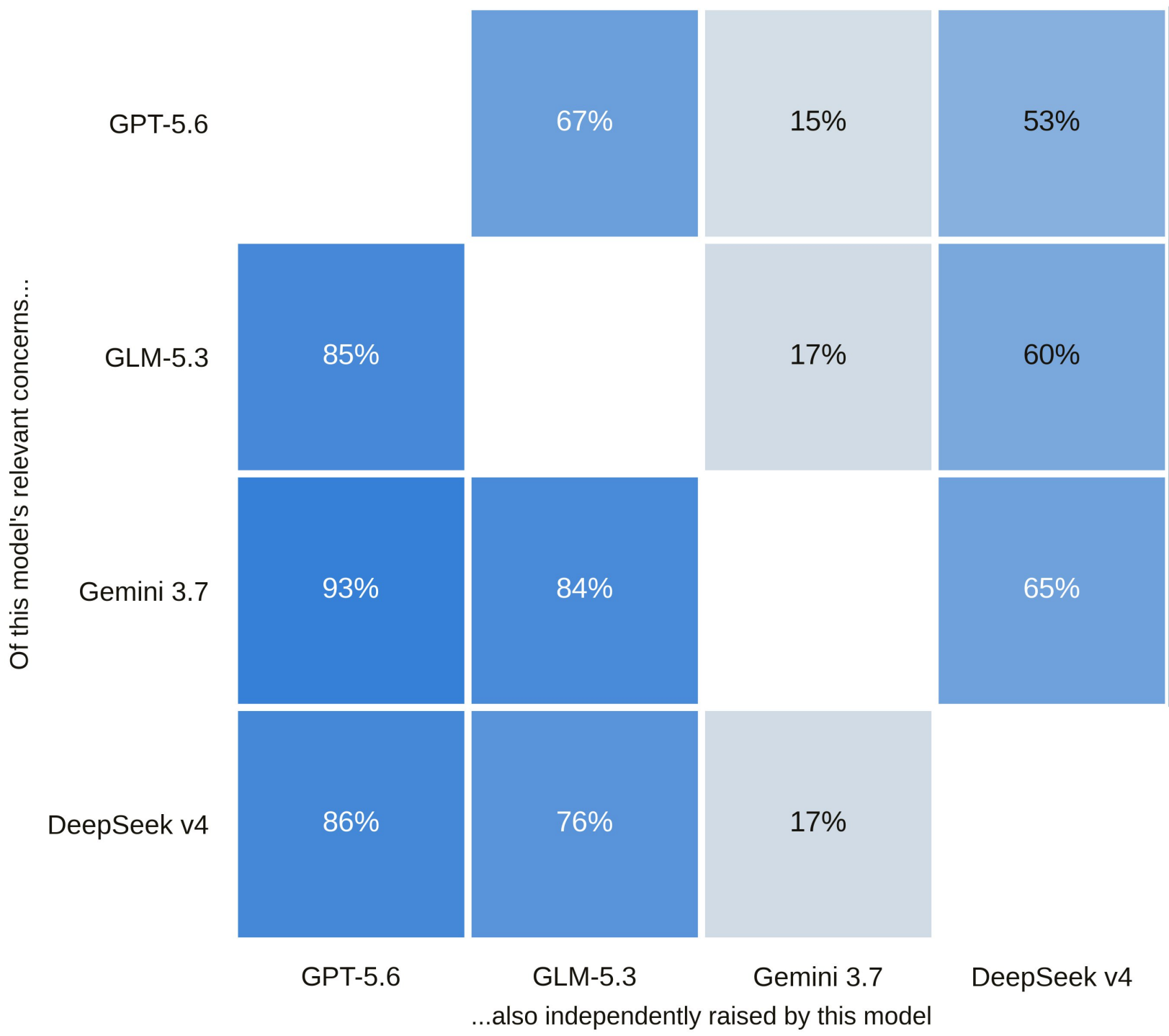

**Figure 7.** Row-normalized overlap in relevant concerns raised between reviewer models. Cell (i,j) is the percentage of row model i's own relevant concerns that column model j also independently raised; the matrix is asymmetric by construction. Own relevant-concern totals: GPT-5.6 Sol 1773, GLM-5.3 Flash 1400, Gemini 3.7 Flash 284, DeepSeek v4 Flash 1099.

## 4 Discussion

Here, we show that reviews produced by four LLMs of articles published in top-tier biomedical journals lead to variable, model-dependent, and generally harsh recommendations. All tested LLMs lacked both sensitivity to errors and specificity, recommending major revision or rejection for papers that had nevertheless been published in top journals.

The main finding of this study is the heterogeneity of the recommendations received for the same manuscript depending on which model is queried. No manuscript that had actually been published was unanimously accepted. The only decision the four models ever agreed on, for 21 of the 200 manuscripts, was major revision. This finding should not be considered specific to LLMs: human reviewers are also known to be highly variable. Committees re-scoring the same submissions in the NeurIPS consistency experiments disagreed on roughly a quarter of accept/reject decisions, with about half of the accepted list changing under a re-run [18, 19], and inter-rater reliability across journal peer review is low by any measure [20, 21]. Here, we show that an algorithmic solution, which carries epistemological hopes of more objective decision-making, is itself similarly variable and fragile, sensitive to both the model and the framing of the context. We do not argue that LLM-generated reviews are no more valid than human reviews – the results reported by Kim and colleagues are actually rather encouraging [13] – but that it is urgent to recognize that LLM assistance in editorial decision-making is dependent on multiple factors, few of which are controllable or transparent. For example, we show that Gemini 3.7 Flash is the most lenient of the tested models, but there is no indication that further versions will follow the same pattern. An editor integrating an LLM into the editorial process would therefore need to continuously monitor the model's behavior on a local benchmark.

We also show a tendency for LLMs to produce harsh reviews, suggesting major revision or rejection for the majority of manuscripts. This behavior likely originates from a shared incentive between models and reviewers to appear authoritative by not accepting a newly submitted manuscript at face value. Trained to be thorough and to appear smart, these models may be optimizing for major revision, the soft middle ground between rejection, which conveys the impression of knowing more than others, and acceptance, which runs against the "fear of missing out" on a great manuscript. From the authors' standpoint, it offers no guarantee, yet warrants additional work.

For a reviewer, a major revision decision can mean two things: the precise identification of one or more problems that can be resolved, or a failure to commit to a decision, which can in turn generate a delay before rejection. Two findings support the latter interpretation for LLMs. First, inserting a blocking methodological error barely changed the decision, even when the error was identified in the review text, suggesting that the decision is not entirely derived from the review comments. Second, framing the submission as a revision significantly shifted the decision toward acceptance, even though the manuscript content remained unchanged. These findings highlight the fragility of editorial recommendations and their bias toward indecision, which may have consequences for researchers.

This distinction also has epistemological implications. A human expert is accountable to the editor who invited them, who can weigh their assessment and choose not to solicit their opinion again. On the other hand, a machine's review, while displaying the same formal apparel and authority, carries something else, even when the concerns themselves are reasonable. Human judgment, expressed in a human's name, should be what calls for human actions – be it additional experiments, invitation to submit elsewhere, or frame the problem differently. We argue that a machine's output can inform that judgment, but it should not substitute for it. The real danger is therefore a shift in function, that a tool built for assistance takes a certification role it has no epistemological legitimacy to occupy. As technology becomes mature and omnipresent, claims that this shift is inevitable will circulate, and are already taking precedent [12].

If the decision itself carries a weak signal, our results nevertheless open the door to collaboration. Indeed, the comments raised by the models themselves, particularly when they converge across several models, appear to contain a signal that could help identify points of vulnerability. Future studies should examine this convergence of signals through expert assessment and determine whether it can provide insights into a manuscript that human reviewers had not detected.

This study has several limitations. First, our gold standard rests on the assumption that papers published in the included journals are of sufficient quality. Retractions do occur, and scientific controversy can discredit a manuscript. However, considering all articles as equally unreliable would plunge research into a relativism that prevents authors and editors from engaging with these issues, and we therefore make the explicit assumption that the editorial work of the journals concerned provides a reasonable basis of trust. Second, none of the reviews underwent expert assessment. Kim and colleagues have already undertaken this substantial work [13] and demonstrated the relative value of these reviews. We are not addressing exactly the same question, instead focusing on LLM recommendations, which have been little explored in the context of biomedical peer review. Third, the LLMs selected were not the most expensive or largest models available. This choice was driven by financial and ecological considerations in this exploratory project, as each experiment carries real consequences in terms of carbon impact. Finally, the open-license articles selected may have leaked into the models' training data. We restricted the corpus to articles published in 2026, but some articles may have been available earlier as preprints, which could have mildly contaminated some experiments – although recognition of an article would more likely have had a positive effect.

### 4.1 Conclusion

In conclusion, our results show that the editorial recommendation made by an LLM is a variable signal, dependent on multiple factors, including model choice and question framing, and should be integrated into the editorial process as such. With new models being released continuously, post-deployment assessment of the behavior of new algorithms against benchmarks that reflect each journal's editorial policies appears essential.

**Data and code availability.** The manuscript corpus (all CC BY 4.0-licensed, §2.1), the flaw-injected manuscript variants produced in Experiment 2 (derivative works of the same CC BY-licensed originals), all review outputs, and the analysis code underlying this preprint will be deposited in a permanent, DOI-assigned public repository prior to journal submission; no licensing restriction on the source material precludes full deposit.

**Author contributions.** B.L.G. conceived the study. B.L.G. and S.H.K. discussed and refined the experimental design. Corpus construction, pipeline implementation, data collection, and analysis were carried out using Claude (Anthropic) as an AI research and coding assistant, directed and reviewed by B.L.G. throughout. B.L.G. and S.H.K. interpreted the results and reviewed the manuscript. R.B. and A.H. contributed to the revision of the original draft.

**Use of AI in this manuscript.** Beyond being the object of study, AI tools (Claude, Anthropic) were used substantially in executing this work: corpus retrieval, experiment implementation, statistical analysis, figure generation, and drafting of this manuscript were performed by an AI agent under the corresponding author's direction and review. The authors take full responsibility for the content, methodology, and conclusions reported.

**Competing interests.** The authors declare no competing interests.

**Funding.** This work received no dedicated funding.

## Appendix A. Exact prompts

All prompts below are reproduced verbatim from the pipeline code, with the sole substitution of `{journal}`, `{category_guidance}` placeholders as noted.

### A.1 Reviewer system prompt (Experiments 1–2), `{journal}` substituted with the manuscript's stated target journal

```
You are a reviewer for {journal}. You will be shown the full text of a submitted
manuscript. Write the review you would hand to the Editor.

Structure your review as:

SUMMARY: <2-3 sentence summary of what the paper claims to show>

MAJOR CONCERNS:
- <substantive issue that would need to be resolved before publication, or
"None.">

MINOR CONCERNS:
- <smaller issues, wording, presentation>

Then end your response with exactly one line in this format, with no other text on
that line:

DECISION: <accept|minor_revision|major_revision|reject>
```

### A.2 Error-injector system prompt (Experiment 2), `{category_guidance}` substituted with one of the two category blocks below

```
You are helping construct a dataset for evaluating peer reviewers' ability to
catch critical flaws. You will be given the full text of a published, presumably
sound paper.

Your task:
1. Identify the paper's central conclusion -- the one claim the abstract and title
are built around.
2. Design one critical flaw to introduce into the paper that, if a reader noticed
it, would undermine that central conclusion. It should fit this general category:

{category_guidance}

   General guidance, not a rigid checklist -- use judgment:
   - It should be a real, substantive problem with the study, specific to this
paper's actual field, data and method -- not a generic template applied on
autopilot.
   - It should be serious enough that a competent reviewer who noticed it would
have to recommend major revision or rejection, not just note it as a minor
concern. If the flaw you're considering would only prompt a passing comment, it's
not the one -- look for something that actually threatens the central conclusion.
   - It should be hard to spot: don't flag, hedge, or apologize for it anywhere in
the text, and don't tip it off with suspicious phrasing. It should just read as a
normal, confidently-written paper.
   - Introduce it with as small an edit as reasonably possible -- typically
replacing or removing a short existing passage (roughly a sentence or few) is
enough; you don't need to touch anything else.

3. Quote the passage you're changing or removing exactly as it appears in the
manuscript below, character for character -- this will be used for an automated
find-and-replace, so it has to match verbatim. Then give the replacement text (or
leave it empty if the passage should just be deleted).

Respond with ONLY a JSON object (no markdown fences, no commentary):
{
  "central_conclusion": "<one sentence: the paper's main claim>",
  "flaw_summary": "<one or two sentences: what the flaw is and why it undermines
the central conclusion>",
  "original_span": "<the exact verbatim passage from the manuscript to change or
remove>",
```

```
  "replacement_span": "<the replacement text, or an empty string if the
original_span should simply be deleted>"
}
```

### Category block: statistical

```
Category: STATISTICAL. Broadly, the analysis applied doesn't really match the
structure of the data or the design -- the kind of thing a careful statistical
reviewer would flag. Use your own judgment about what fits this paper; a few loose
examples of the shape this can take (don't feel bound by them, and don't reuse
them verbatim): an analysis that assumes independent observations when the data
are clustered or repeated; multiple testing with no correction or disclosure;
ignoring censoring; unmodeled confounding or stratification. Pick whatever's most
natural and well-hidden for this specific paper's actual data and analysis.
```

### Category block: methodological

```
Category: METHODOLOGICAL. Broadly, either the outcome was produced by someone or
something with undisclosed access to information that could bias the judgment, or
the instrument/reagent/measurement doesn't quite isolate what's claimed. Use your
own judgment about what fits this paper; a few loose examples of the shape this
can take (don't feel bound by them, and don't reuse them verbatim): an unblinded
assessor scoring a subjective outcome; ground truth produced by someone who'd
already seen the result being evaluated; a reagent or instrument with a known
specificity/validity limitation used without disclosing it. Pick whatever's most
natural and well-hidden for this specific paper's actual method.
```

## A.3 Detection-judge system prompt (Experiment 2)

```
You are checking whether a peer review caught a specific planted flaw in a
manuscript. You will be given: the paper's real central conclusion, a description
of the one critical flaw that was planted, and the full text of a review someone
wrote of the doctored manuscript.

Decide whether the review explicitly identifies this specific flaw -- the actual
methodological/logical problem described, in the same location -- as opposed to
just raising generic or unrelated criticisms that happen to be in the
neighborhood.

Respond with ONLY one word: yes, no, or partial.
```

## A.4 Review-quality judge system prompt (Experiment 3)

```
You are an experienced journal editor evaluating four peer reviews of the same
submitted manuscript, to assess review quality. You do not know who wrote each
review -- treat all four as anonymous submissions labeled Review A, Review B,
Review C, and Review D, in no particular order, and judge them purely on their
content.

Read the manuscript and all four reviews carefully, then:

1. Identify every major concern raised across all four reviews -- a substantive
methodological, statistical, or interpretive objection, not a typo, formatting
note, or citation request. Group concerns that are substantively the same
underlying issue, even if phrased differently or focused on a different symptom of
it, into a single cluster, and list every review letter that raised it.

2. For each concern cluster, classify it as "relevant" (a real, well-founded issue
that could legitimately affect the paper's conclusions or its acceptance) or
"nitpicking" (technically a criticism, but minor, pedantic, or unlikely to matter
-- e.g. requesting an additional citation, quibbling over word choice, or a point
the manuscript already addresses elsewhere).

3. Rate each review's overall constructiveness on a 1-5 scale: 1 = purely critical
with no actionable guidance for the authors; 5 = highly constructive, specific,
and clearly aimed at helping the authors improve the paper (independent of how
harsh or lenient its overall recommendation is).

4. Pick the single best review of the four -- the one you would most want an
author to receive -- and say why in one sentence.

Respond with ONLY a JSON object (no markdown fences, no commentary):
{
```

```
  "concern_clusters": [
    {"summary": "<short phrase describing the concern>", "raised_by": ["A", "C"],
"category": "relevant"},
    {"summary": "<short phrase describing the concern>", "raised_by": ["B"],
"category": "nitpicking"}
  ],
  "constructiveness": {"A": <1-5>, "B": <1-5>, "C": <1-5>, "D": <1-5>},
  "best_review": "A",
  "best_review_reason": "<one sentence>"
}
```